\documentclass{iau}
\usepackage{graphicx}

\title[Numerical study of AGN feedback] %% give here short title %%
{Numerical study of active galactic nucleus feedback in an elliptical galaxy with {\it MACER}}

\author[Feng Yuan et al.]   %% give here short author list %%
{Feng Yuan$^1$, Jeremiah P. Ostriker$^2$,  DooSoo Yoon$^1$,  Ya-Ping Li$^{1}$,  Luca Ciotti$^3$, Zhao-Ming Gan$^{1}$,  Luis C. Ho$^{4,5}$, and Fulai Guo$^1$}

\affiliation{$^1$Shanghai Astronomical Observatory, Chinese Academy of Sciences, \\
80 Nandan Road, Shanghai 200030, China; email: {\tt
fyuan@shao.ac.cn}
\\[\affilskip]
$^2$ Department of Astronomy, Columbia University, 550 W. 120th Street, New York,  USA
\\[\affilskip]
$^3$ Department of Physics and Astronomy, University of Bologna,  40129 Bologna, Italy
\\[\affilskip]
$^4$ Kavli Institute for Astronomy and Astrophysics, Peking University, Beijing 100871, China
\\[\affilskip]
$^5$ Department of Astronomy, School of Physics, Peking University, Beijing 100871, China}

\pubyear{2018}
\volume{342}  %% insert here IAU Symposium No.
 \jname{Perseus in Sicily: from black hole to
cluster outskirts} \editors{Keiichi Asada, Elisabete de Gouveia dal Pino, Hiroshi Nagai, Rodrigo Nemmen, \& Marcello Giroletti, eds.}
\begin{document}

\maketitle

\begin{abstract}
   This paper summarizes our recent works of studying AGN feedback in an isolated elliptical galaxy by performing high-resolution hydrodynamical numerical simulations. Bondi radius is resolved and the mass accretion rate of the black hole is calculated. The most updated AGN physics, namely the discrimination of  cold and hot accretion modes and the exact descriptions of the AGN  radiation and wind for a given accretion rate are adopted and their interaction with the gas in the host galaxy is calculated. Physical processes such as star formation and SNe feedback are taken into account. Consistent with observation, we find the AGN spends most of the time in the low-luminosity regime.  AGN feedback overall suppresses the star formation; but depending on location in the galaxy and time, it can also enhance it. The light curve of specific star formation rate is not synchronous with the AGN light curve. We find that wind usually plays a dominant role in controlling the AGN luminosity and star formation, but radiation also  cannot be neglected.
\keywords{accretion, accretion disks -- black hole physics --  galaxies: evolution -- galaxies: nuclei}
\end{abstract}

\firstsection % if your document starts with a section,
\section{Introduction}

There is growing evidence for the coevolution of central
supermassive black holes and their host galaxies (see review by Kormendy \&
Ho 2013). It is generally believed that active galactic nucleus
(AGN) feedback plays an important role in the evolution of
galaxies from both observational and theoretical arguments (see review by Fabian 2012; King \& Pounds 2015; Naab \& Ostriker 2017).  The basic scenario of the AGN feedback is illustrated by Figure 1. The outputs from the AGN, namely radiation, wind, and jet, interact with the interstellar medium (ISM) in the host galaxy and  change its density and temperature. Consequently,  star formation rate will be changed, which changes the evolution of the galaxy. On the other hand, the change of the gas properties will in turn  affect the fueling of the AGN. Obviously, the most crucial  factor to determining the effects of AGN feedback is the outputs from AGN, which is determined by  the value of AGN accretion rate and  the AGN outputs for  a given accretion rate.

There have been many  works on AGN feedback using hydrodynamical numerical simulation. Most of these work focus on very large spatial scales, their resolution is typically several $kpc$ or even larger. Thus it is difficult to resolve the Bondi radius, which is typically several tens of $pc$. In this case, they have to estimate the accretion rate, which may have an uncertainty as large as $\sim 300$ (Negri \& Volonteri 2017; see also Korol et al. 2016 for an analytical discussion).  Moreover, although after several decades' effort we now have quite robust understanding to accretion physics, these physics unfortunately have not been  properly adopted in most current  works of AGN feedback. Lastly, the interaction of the AGN outputs and the ISM is also not carefully calculated.

In our works,  we study AGN feedback by focusing on a single galaxy using our  high-resolution two-dimensional numerical simulations. There are three key features in our model. First, the inner boundary of the simulation domain is ten times smaller than Bondi radius so we can precisely calculate the accretion rate. Second, we adopt the most updated AGN physics. Third, we carefully calculate the interaction between AGN outputs and ISM.

\begin{figure}
% \vspace*{-2.0 cm}
\begin{center}
 \includegraphics[width=2.5in]{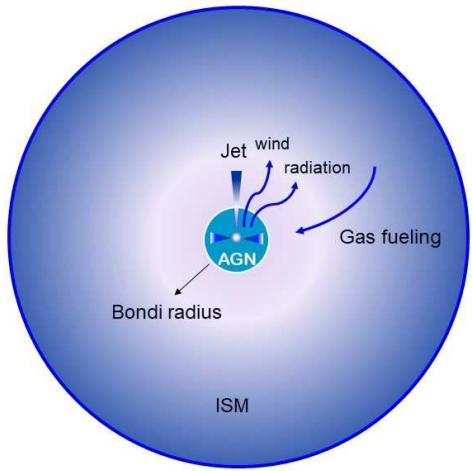}

% \vspace*{-1.0 cm}
 \caption{Schematic figure of AGN feedback. }
   \label{fig2}
\end{center}
\end{figure}

\section{AGN accretion physics}

Since the inner boundary of our simulation domain ($r_{\rm in}$) is $\sim 0.1 R_{\rm Bondi}$,  we do not need to use the Bondi formula, which is an approximation, to calculate the accretion rate.  Instead, the accretion rate at $r_{\rm in}$  can be directly calculated by the  following  precise equation (we use spherical coordinates):
\begin{equation}
     \dot{M}(r_{\rm in}) = 2\pi r_{\rm in}^2\int^{\pi}_0 \rho(r_{\rm in},\theta) ~ {\rm min} \left[v_{r}(r_{\rm in},\theta),0\right] \sin{\theta}\,d\theta.
\label{mdotbondi}
\end{equation}
According to the black hole accretion theory and observations of black hole X-ray binaries (Yuan \& Narayan 2014 and references therein),  black hole accretion is divided into ``cold'' and ``hot'' modes. The boundary between them is $\dot{M}\approx 2\%\dot{M}_{\rm Edd}(\equiv 20\%L_{\rm Edd}/c^2)$ assuming a $10\%$ radiative efficency.  In the following, we  describe radiation and wind  in each mode. Jet is neglected since we assume it deposits very little energy in the galaxy. It is necessary to examine this assumption in  future work.

\subsection{Cold accretion mode}
Within $r_{\rm in}$, the accretion still cannot be resolved and must be treated as subgrid physics. The infalling gas initially will freely fall until a small disk is formed. So there will be a time lag and difference of values between the accretion rate in the small disk ($\dot{M}_{\rm BH}$) and $\dot{M}(r_{\rm in})$ (see Yuan et al. 2018 for details).  The luminosity emitted from the disk is $L_{\rm BH}=\epsilon_{\rm cold}\dot{M}_{\rm BH}c^2$. The corresponding Compton temperature needed to calculate the Compton heating is $T_{\rm C,cold}=2\times 10^7 {\rm K}$.

Winds in the cold mode have been widely observed in many sources, so we have good observational constraint on the properties of wind.  In our work, we adopt the results compiled in Gofford et al. (2015). The mass flux and velocity of wind are described by:
\begin{equation}
 \dot{M}_{\rm W,C} = 0.28\,  \,\left( \frac{L_{\rm BH}}{10^{45}\,\rm erg\,s^{-1}} \right)^{0.85}M_{\odot}\,{\rm yr^{-1}},
\end{equation}
\begin{equation}
    v_{\rm W,C} = 2.5\times10^{4}\, \,\left( \frac{L_{\rm BH}}{10^{45}\,\rm erg\,s^{-1}} \right)^{0.4}{\rm km\,s^{-1}}.
\label{coldwindvelocity}
\end{equation}
In Gofford et al. (2015), the winds are detected at a distance of
$10^{2-4}r_s$ from the black hole, which is roughly consistent with the value of the inner boundary of our simulation.

\subsection{Hot accretion mode}
In the hot accretion mode, the accretion flow consists of an outer truncated thin disk and an inner hot accretion flow. The truncation radius depends on the mass accretion rate (Yuan \& Narayan 2014),
 \begin{equation}
    r_{\rm tr} = 3r_s\left[\frac{2\times 10^{-2}\dot{M}_{\rm Edd}}{\dot{M}(r_{\rm in})}\right]^2.
 \end{equation}
The dynamics and radiation of the hot accretion flow have been intensively studied and are well understood (see review by Yuan \& Narayan 2014).  Different from the thin disk, the radiative efficiency of a hot accretion flow is a function of accretion rate (Xie \& Yuan 2012):
\begin{equation}
\epsilon_{\rm EM,hot}(\dot{M}_{\rm BH})=\epsilon_0\left(\frac{\dot{M}_{\rm BH}}{0.1L_{\rm Edd}/c^2}\right)^a,
\end{equation}
the values of $\epsilon_0$ and $a$ are given in Xie \&
Yuan (2012). Note that the efficiency is comparable to that of the thin disk when the accretion rate is high. The radiation emitted from a hot accretion flow has relatively more hard photons compared to a thin disk; thus, the Compton temperature is higher, $T_{\rm C,hot}\approx 10^8{\rm K}$ (Xie, Yuan \& Ho 2017).

For the wind in the hot accretion mode,  although we are accumulating more and more observational evidences (e.g., Wang et al. 2013; Cheung et al. 2017), most of them are indirect evidences, and so it is hard to directly get the observational constraint on the properties of wind. By contrast,  we have very good theoretical understanding to the wind from hot mode (see review in Yuan et al. 2015; Yuan \& Narayan 2014).  In particular,  the mass flux and velocity  of wind have been well studied based on three-dimensional general relativity MHD numerical simulations (Yuan et al. 2015),
\begin{equation}
    \dot{M}_{\rm W,H}\approx \dot{M}(r_{\rm in})\left[1-\left(\frac{3r_s}{r_{\rm tr}}\right)^{0.5}\right],
\label{hotwindflux}
\end{equation}
\begin{equation}
v_{\rm W,H}\approx (0.2-0.4) v_{\rm K}(r_{\rm tr}).
\label{windvelocity}
\end{equation}

\section{Model}
The evolution of the galactic gas flow, with the effect of AGN feedback, is described by the conservation of mass, momentum, and energy (e.g., Ciotti \& Ostriker 2012; Ciotti et al. 2017; Pellegrini et al. 2018).  We solve them by the parallel ZEUS code using two-dimensional axisymmetric spherical coordinates. The radial direction of simulation domain covers the range of
2.5 pc -- 250 kpc, the finest resolution is at
the innermost grid, $\sim$0.3 pc. The wind and radiation are injected at the inner boundary, and their interaction with ISM is calculated. Other physical processes such as the radiative cooling, star formation, and SNe are taken into account. We name our code \textit{MACER} (\textit{M}assive \textit{A}GN \textit{C}ontrolled \textit{E}llipticals \textit{R}esolved).

We have investigated two cases, with the specific angular momentum of the has being low and high in Yuan et al. (2018) and Yoon et al. (2018), respectively. In this paper, we will take Yuan et al. (2018) as an example when we introduce the simulation results.

\section{Results}
We have simulated several models: noFB (no AGN feedback included), fullFB (both wind and radiation feedback included), windFB (only wind), and radFB (only radiation). The main results are summarized below.

\subsection{AGN light curve}

Figure 2 shows the AGN light curves.  For the noFB model, the light curve is featureless; it gradually decreases  because of the depletion of gas in star formation. Once feedback is included,  the AGN luminosity strongly fluctuates. This is because, when the AGN is luminous, strong radiation and wind will be emitted, pushing the surrounding gas away and heating it. Subsequently, the accretion rate of the AGN will substantially decrease, and the AGN will dim. The surrounding gas will then gradually cool by radiation and become dense, and so the accretion rate will increase. From the figure, we can see that the AGN spends most of its time in the low-luminosity phase (hot mode), with a typical luminosity $L\sim 10^{-4}L_{\rm Edd}$.

The variability amplitudes for the radFB and windFB models are similar, and both of
them are similar to the fullFB model. This indicates that
wind and radiation feedback  cause a similar amplitude of
variability. However, in the time-average sense,  the AGN luminosity in the radFB
model is $10^{-2}L_{\rm Edd}$,  almost two orders of magnitude higher
than that in the windFB model. The main reason for such a difference is that wind can deposit their momentum much more efficiently than radiation (Yuan et al. 2018).

We can see that in the fullFB model, the ¡°baseline¡± AGN luminosity is very similar to that of the
windFB model. This indicates that the mass accretion rate of
the black hole is controlled by the wind feedback rather than
the radiation. However, when we carefully compare the zoom-in plots of the fullFB
and windFB models, we find that there are  more outbursts in the fullFB model than in the
windFB model (Yuan et al. 2018). This indicates that feedback by radiation and
wind are coupled together, and neither of them can be
neglected. At last, the light curve amplitude suddenly decreases after $\sim$ 8 Gyr. The large amplitude is mainly because of the strong ``perturbation'' of the strong wind in  the cold feedback mode.  Due to the gradual loss of the gas in the galaxy, after $\sim 8$ Gyr, the AGN fails to reach the cold accretion mode and the strong wind disappears.

\begin{figure*}
%[!htbp]
%[!htbp]
%\centering
\begin{center}
   \includegraphics[width=1.0\textwidth]{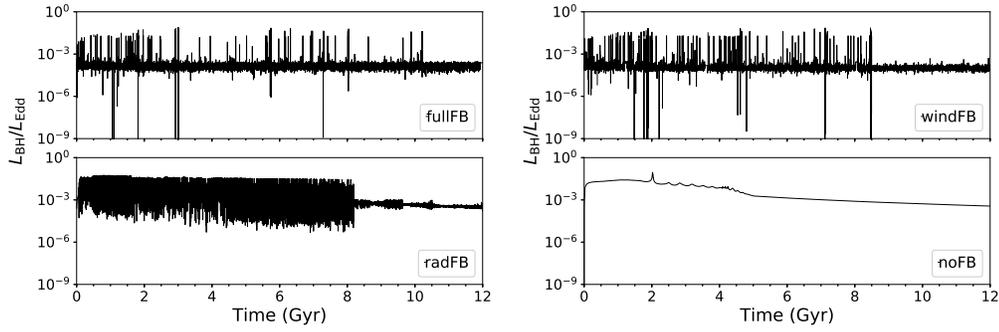}
       \vspace{0.1cm}
    \caption{Light curves of AGN luminosity as a function of time for various models.  For clarity, we choose the data point so that the two adjacent ones have a relatively large time interval of 2.5 Myr; in this case, some outbursts are filtered out. Taken from Yuan et al. (2018). }
    \end{center}
\end{figure*}

We have also calculated the AGN duration (lifetime) using our simulation data. We find that the typical value is $\sim 10^5$ yr. This is consistent with some observations (e.g., Schawinski et al. 2015). We note that the typical lifetime obtained in Gan et al. (2014), which has the same model framework but different AGN physics, is almost two orders of magnitude higher. This indicates the impact of specific AGN physics adopted in the feedback model.

\subsection{Mass growth of the black hole}

We have calculated the growth of the black hole mass. When no AGN feedback is included, we confirm (see e.g., Ciotti et al. 2017) that the  mass can easily reach above $10^{10}M_{\odot}$, which is obviously too large. In the radFB model, interestingly, we find that the black hole mass becomes even larger. This is because when we include radiation, the star formation, which can deplete some gas in the galaxy, becomes weaker due to the radiative heating. Therefore there will be more gas left to fuel the black hole. In the windFB model, the final black hole mass becomes substantially smaller, $\sim 2\times 10^9 M_{\odot}$, only slightly larger than the initial value. This indicates that wind plays a dominant role in suppressing the accretion rate and  the black hole mass. Finally, for the fullFB model, the black hole mass is slightly higher than that in the windFB model. The reason is that star formation becomes weaker when radiative feedback is included.

\subsection{Star formation}

\begin{figure*}[!htbp]
    \begin{center}$
        \begin{array}{ccc}
            \includegraphics[width=0.49\textwidth]{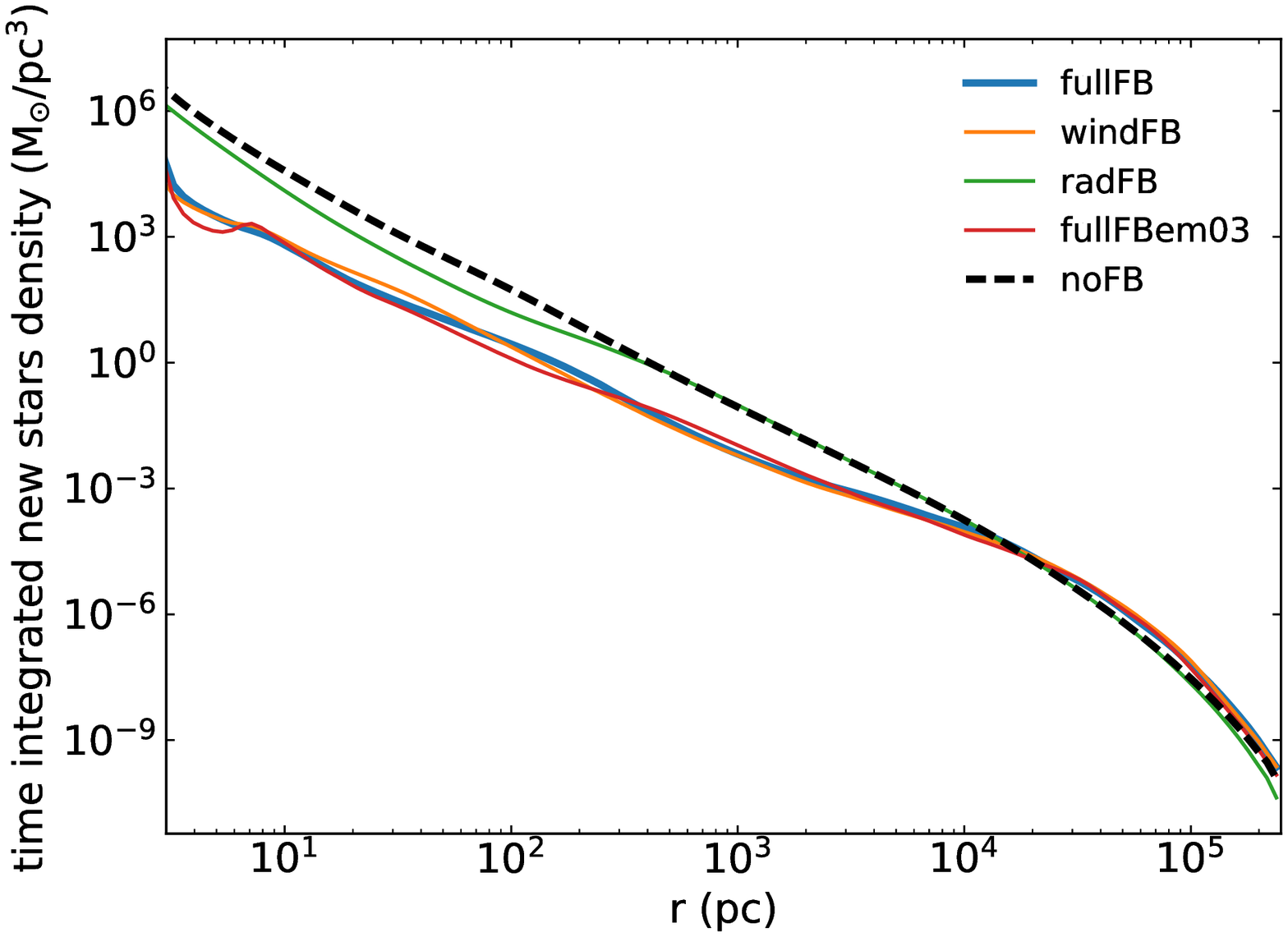} &
            \includegraphics[width=0.49\textwidth]{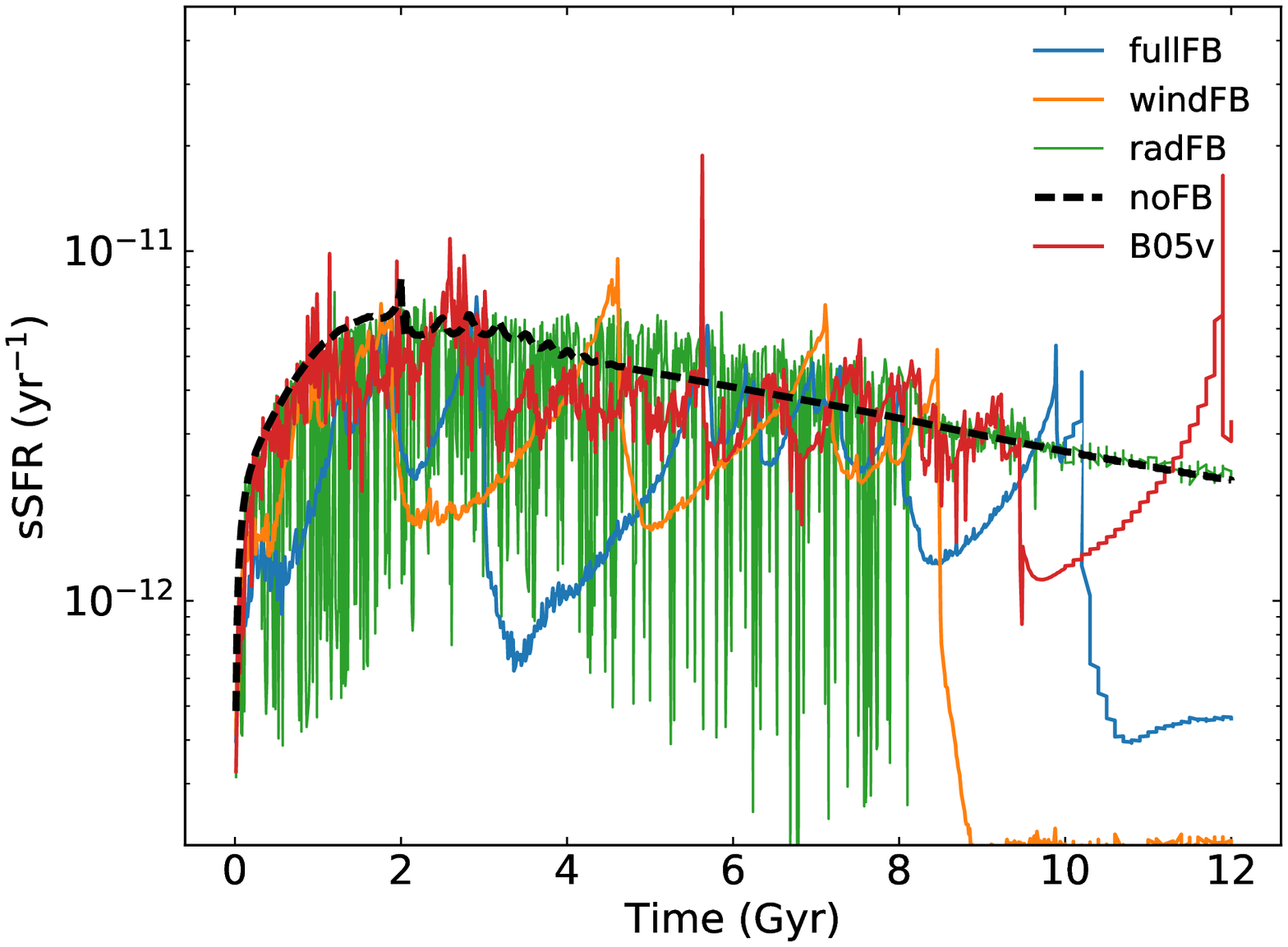}
       \end{array}$
    \end{center}
    \caption{Effects of AGN feedback on star formation for various models. Left panel: Time-integrated  mass of newly born stars at a given radius per unit volume. Right panel: Specific star formation rate over time for various models. Taken from Yuan et al. (2018).}
    \label{fig:newst}
\end{figure*}
The left plot of Figure 3 shows the time-integrated total mass of
newly born stars per unit volume as a function of radius. In the radFB model, within several hundred pc, star formation is slightly suppressed compared to noFB model.  In the windFB model, we can see that star formation is strongly suppressed all the way up to $\sim 20 {\rm kpc}$. This is  because of the momentum feedback of the wind, i.e., winds push the gas away from the central region beyond $\sim 20 {\rm kpc}$. We note that the
Gemini integral field unit observations by Liu et al. (2013) do find the wind can reach that distance.  The result of fullFB model is very similar to windFB, indicating that wind plays a dominant role in controlling star formation. At the region of $r\gtrsim 20  {\rm kpc}$,  the gas is accumulated there, so star formation is slightly enhanced.

The right plot of Fig. 3 shows the evolution of specific star formation rate (sSFR), i.e., the star formation rate normalized by the stellar mass of the galaxy. We find three important results. First, by comparing the windFB and fullFB models, we can see that their general patterns are similar, but their light curves are not synchronous with each other. There is
an obvious offset between them, and the ``amplitudes'' of the
fullFB model are also larger. This indicates that, although in
the time-integrated sense the wind seems to be much more
important than radiation in suppressing star formation, radiation also plays a very
important role. The wind and radiation couple together in
affecting star formation. Second, by comparing the fullFB and noFB models, we can see that the sSFR in fullFB model is in general  suppressed compared to the noFB model, but
occasionally the sSFR  can also be enhanced. Third, by comparing the light curve of sSFR of the fullFB model with the AGN light curve shown in Fig. 2,  we find that they are not synchronous with each other. This is partly because the timescale of star formation episodes ($\tau_{SF}\gtrsim 100 {\rm Myr}$) is much longer than the timescale of AGN activity ($\tau_{\rm AGN}\lesssim 1 {\rm Myr}$) (e.g., Harrison 2017). The last two results put a serious challenge to the observational test of the effect of AGN feedback on star formation, and this may explain why the conclusion reached by observational studies of the relation between AGN and star formation is so diverse (see review by Xue 2017).

\subsection{AGN duty cycle}

From the AGN light curve, we can calculate the AGN duty cycle, which is defined as the percentage of the total time of AGN spent below or above a given Eddington ratio. From the light curve shown in Fig. 2, we can see that the AGN must spend  most of the time in the low-luminosity regime, i.e, the hot accretion (feedback) mode. In fact, we find that
AGN spends over 80\% of its evolution time with Eddington
ratios below $2\times 10^{-4}$.  We have compared the simulation result with the observational data which are available for low-redshift sources and general consistency is found. These results suggest the potential importance of feedback effect by low-luminosity AGNs.

Observations show that, although AGNs spend most of their time in the low-luminosity
AGN phase,  they  emit most of their energy during the
high-luminosity phase. To compare with this observation, we have  calculated the percentage of the total energy
emitted above or below a given Eddington ratio. Unfortunately, we find that the AGN only emits 6\% of the entire energy at the Eddington ratio above 0.02,  which is not consistent with observation. What is the reason? Note that in our current model  we only adopt one value for each model parameter and have not done any parameter survey. Our ongoing work of studying the effect of different model parameter seems to indicate that this discrepancy with observation can be  solved by adjusting the values of some parameters (Yao et al. in preparation).

\end{document}